\documentclass[english]{article}
\usepackage{amsfonts}
\usepackage{amsmath}
\usepackage{amssymb}
\usepackage{amsthm}
\usepackage{verbatim}
\usepackage{float}
\usepackage{subfig}
\usepackage{multirow}
\usepackage[colorlinks,citecolor=blue,urlcolor=blue,linkcolor=red]{hyperref}
\usepackage{fullpage}
\usepackage[affil-it]{authblk}
\usepackage{siunitx}
\usepackage{textcomp}
 \usepackage{tabularx}
\usepackage{booktabs}
\usepackage{caption}
\usepackage{graphicx}

\renewcommand{\phi}{\varphi}
\renewcommand{\epsilon}{\varepsilon}

\begin{document}
\title{Deep Learning Algorithms to Isolate and Quantify the Structures of the Anterior Segment in Optical Coherence Tomography Images}

\author[1,2]{Tan Hung Pham}
\author[1]{Sripad Krishna Devalla}
\author[3]{Aloysius Ang}
\author[2]{Soh Zhi Da}
\author[4]{Alexandre H. Thi\'{e}ry}
\author[1,5,6]{Craig Boote}
\author[2 $\star$]{Ching-Yu Cheng}
\author[7 $\star$]{Victor Koh}
\author[1,2 $\star$]{Micha\"el J. A. Girard}

\affil[1]{Ophthalmic Engineering and Innovation Laboratory, Department of Biomedical Engineering, Faculty of Engineering, National University of Singapore, Singapore.}
\affil[2]{Singapore Eye Research Institute, Singapore National Eye Centre, Singapore.}
\affil[3]{Yong Loo Lin School of Medicine, National University of Singapore, Singapore}
\affil[4]{Department of Statistics and Applied Probability, National University of Singapore, Singapore.}
\affil[5]{School of Optometry and Vision Sciences, Cardiff University, UK.}
\affil[6]{Newcastle Research and Innovation Institute, Singapore.}
\affil[7]{Department of Ophthalmology, National University Hospital, Singapore.}

\bigskip

\affil[$\star$]{All three contributed equally and all three are corresponding authors.}

\maketitle


%
%
\begin{abstract}
\noindent

\textbf{}
Accurate isolation and quantification of intraocular dimensions in the anterior segment (AS) of the eye using optical coherence tomography (OCT) images is important in the diagnosis and treatment of many eye diseases, especially angle closure glaucoma. In this study, we developed a deep convolutional neural network (DCNN) for the localization of the scleral spur, and the segmentation of anterior segment structures (iris, corneo-sclera shell, anterior chamber). With limited training data, the DCNN was able to detect the scleral spur on unseen ASOCT images as accurately as an experienced ophthalmologist; and simultaneously isolated the anterior segment structures with a Dice coefficient of 95.7\%. We then automatically extracted eight clinically relevant ASOCT parameters and proposed an automated quality check process that asserts the reliability of these parameters. When combined with an OCT machine capable of imaging multiple radial sections, the algorithms can provide a more complete objective assessment. This is an essential step toward providing a robust automated framework for reliable quantification of ASOCT scans, for applications in the diagnosis and management of angle closure glaucoma.

\end{abstract}

\section{Introduction}
\label{sec.intro}
By 2020, the number of people affected by primary angle closure glaucoma (PACG) is estimated to be up to 23.4 million\cite{RN1,RN10}. PACG is associated with a high rate of blindness\cite{RN34,RN35} that is up to 5 times greater than primary open-angle glaucoma\cite{RN26}. Therefore, an early diagnosis followed by effective management strategies is essential to reduce the damage to the optic nerve head tissues that could lead to irreversible vision loss\cite{RN9}. Early diagnosis is crucial in the Asian population, given the higher prevalence of PACG compared to European and African populations\cite{RN34,RN35,RN36}. \\

\begin{figure}[H]
    \centering
    \includegraphics[scale=0.7]{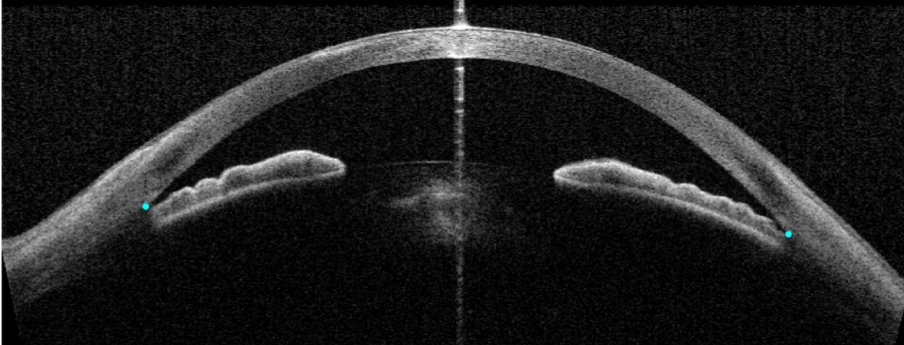}
    \caption{ Example of scleral spur location (cyan dots) on a well-centered anterior segment optical coherence tomography scan.}
    \label{suppli:7}
\end{figure}

The diagnosis of PACG is based on the status of the anterior chamber angle (ACA)\cite{RN2,RN11,RN12}. While the gold standard for ACA assessment is dark-room indentation gonioscopy\cite{RN37}, the procedure requires direct contact with the eye and is highly dependent on the physician’s expertise and the background illumination\cite{RN37,RN3}. This can result in poor reproducibility and diagnostic accuracy. In contrast, anterior segment optical coherence tomography (ASOCT) imaging allows for an objective, fast and non-contact assessment of the ACA in a standardized dark-room environment\cite{RN3,RN38}. However, current technology typically requires the manual identification and marking of the scleral spur location (SSL) (\textbf{Figure \ref{suppli:7}}) by a human grader before ACA measurements such as trabecular iris space area (TISA) and angle opening distance (AOD) can be measured to quantify the anterior chamber angle\cite{RN4}. The introduction of this subjective human factor has been shown to introduce significant intra- and inter-observer variability\cite{RN3,RN38,RN4}. The inconsistent labelling of SSL compromises the diagnosis and the monitoring of treatment effectiveness/disease severity in PACG\cite{RN4}. In addition, with swept-source ASOCT imaging, there are up to 128 cross-sectional scans obtained per eye. To manually label each individual section in a timely manner would not be clinically viable, and therefore automated image processing algorithms are required.\\

Deep convolutional neural networks (DCNNs) have been shown to perform well with many medical imaging modalities\cite{RN23,RN24,RN18,RN31,RN30}, but their applications in ASOCT imaging are  nascent. From the perspective of the current study, there are two relevant applications that can benefit from DCNNs, namely: object localization (for SSL detection) and segmentation (for classifying tissues such as the cornea and the iris). Traditional object detection and localization approaches in DCNNs are mainly based on classification and regression\cite{RN5}. However, this approach requires a large number of labelled images to achieve robust automation\cite{RN6}. Moreover, accurate landmark localization is critical for the diagnosis and management of PACG. Hence with limited training data, a traditional regression approach is not ideal in providing a high accuracy prediction. Frequently, in the medical context, it might not be feasible to obtain a large number of labelled images due to limited resources and time. This problem is exacerbated in certain ocular conditions that are relatively less common which may benefit from mass screening such as PACG. In addition, the reduced availability of ASOCT images for eyes with PACG can be attributed to the lack of accessible equipment, cost, and clinical expertise. \\

In this study, we developed a custom hybrid DCNN inspired from widely used U-Net and full-resolution residual network (FRRnet)\cite{RN20} for the localization of scleral spur, and the segmentation of the anterior segment structures (iris, corneo-sclera shell, anterior chamber). The hybrid DCNN leveraged the U-Net architecture to simultaneously exploit the local (i.e. tissue texture) and contextual (i.e. tissue spatial arrangement) information and exploited the FRRnet pathway to achieve precise localization. Further, we automatically extracted eight clinically relevant ASOCT parameters from the segmented structures. The aim of the work is to offer a robust and automated framework for the accurate localization of the scleral spur and quantification of the ASOCT structures for enhancing the diagnosis and management of PACG.\\

\section{Methods}
\subsection{ASOCT imaging}
\begin{figure}[H]
    \centering
    \includegraphics{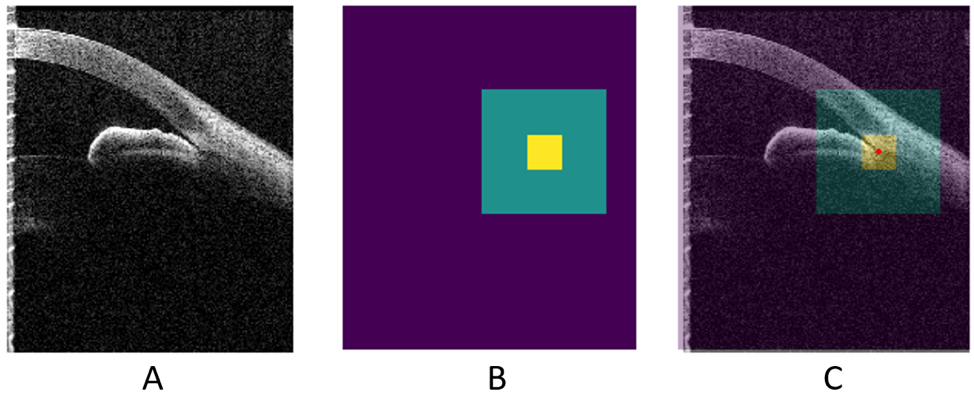}
    \caption{ Labelled data for the SSL. \textbf{(A)} Input ASOCT image; \textbf{(B)} ground truth of the SSL (yellow represents focus region, cyan represents attention region) and \textbf{(C)} prediction of the SSL with red dot as the center of focus region.}
    \label{fig:8}
\end{figure}

We included ASOCT images from patients examined at the Eye Surgery Centre, National University Hospital, Singapore. Prior informed consent was obtained for all patients. The study was conducted in accordance with the tenets of the World Medical Association’s Declaration of Helsinki and had ethics approval from the National Healthcare Group Domain Specific Review Board. In total, ASOCT images from 100 patients (175 eyes) were included for analysis. The scans were obtained from the swept-source Casia SS-1000 ASOCT (Tomey Corporation, Nayoga, Japan). For each eye, a 360-degree scan yielded up to 128 cross-sections of the anterior segment. We used 620 images from 42 patients (75 eyes) for training and another 200 images from a further 58 patients (100 eyes) for testing. Since each image contained two scleral spur instances, we further divided the images in half for scleral spur localization (\textbf{Figure \ref{fig:8}}).

\subsection{Small landmark localization and ASOCT segmentation}

The accurate localization of small landmark points using neural networks has always been challenging\cite{RN61}. In the current study, we adopted a segmentation approach for both the landmark localization and the ASOCT segmentation. A MATLAB (R2018a, MathWorks Inc., Natick, MA) script was prepared to assist in labelling the SSL (landmark localization). Three definitions were used to locate the scleral spur: \textbf{1)} A change in curvature in the corneo-scleral interface; \textbf{2)} The posterior end of the trabecular meshwork; and \textbf{3)} The posterior end of a protruding structure along the cornea and sclera\cite{RN4,RN27}. In each image, the following classes were identified (\textbf{Figure \ref{fig:8}}): \textbf{(1)} focus region (in yellow); \textbf{(2)} attention region (in cyan); and \textbf{(3)} the background (in purple).\\

\begin{figure}[H]
    \centering
    \includegraphics[scale=0.5]{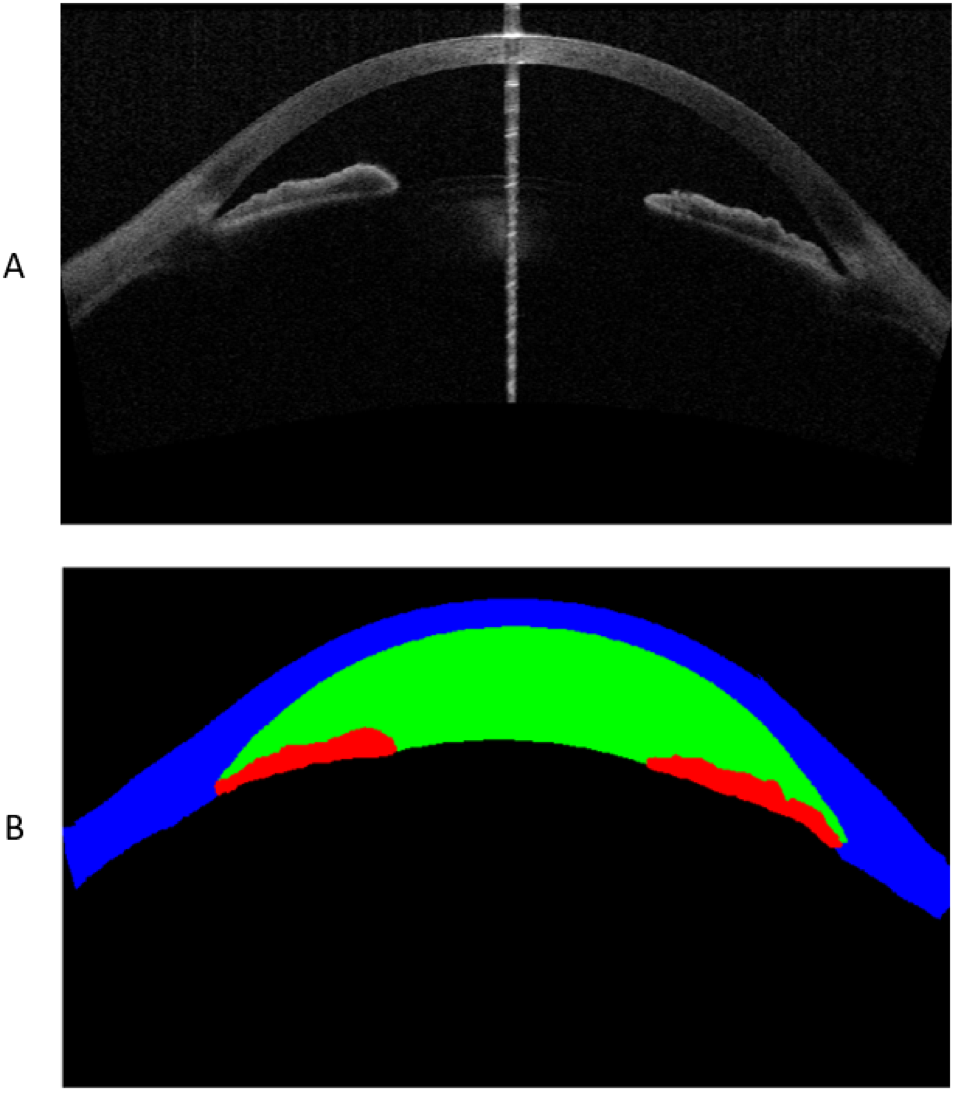}
    \caption{ Labelled data for ASOCT segmentation. \textbf{A}: input image. \textbf{B}: ground truth for output, red: iris, blue: corneo-sclera shell, green: anterior chamber.}
    \label{fig:9}
\end{figure}

FIJI\cite{RN56} was used to obtain the manual segmentations of the ASOCT tissues. In each image, the following classes were identified (\textbf{Figure \ref{fig:9}}): \textbf{(1)} the iris (in red);\textbf{(2)} the corneo-sclera shell (in blue); \textbf{(3)} the anterior chamber (in green); and the background (in black). The SSL labelling and the manual segmentations used for training the DCNNs were prepared by two trainers: a trained medical student (AA), and a trained observer (THP), both with more than two years of experience in ASOCT imaging.\\

The landmark localization and segmentation performance of the DCNNs on unseen ASOCT images were evaluated by three graders: the aforementioned trained observer (observer A; THP) and medical student (observer B; AA), and a glaucoma fellowship trained ophthalmologist (observer C; VK) with eight years of experience in the management of PACG. \\

\subsection{Measurement of ASOCT parameters}

The ASOCT parameters could be automatically measured once the scleral spur was defined and the anterior segment intraocular tissues segmented. The key structural parameters, including ACA, anterior chamber and iris-based measurements were automatically computed based on their definitions \textbf{Table \ref{tab:table1}}.

\subsection{Network architecture and training}
In recent years, several research groups have successfully used U-Net and its variants\cite{RN29,RN30,RN28,RN18} in medical image segmentation. The sequential downsampling and upsampling of images combined with skip connections\cite{RN57} help in simultaneously extracting both the local (i.e., tissue texture) and contextual (i.e., tissue spatial arrangement) information. This allows U-Net style architectures to achieve very high levels of segmentation accuracy even when trained with limited training data\cite{RN18,RN30}. Another promising but less explored DCNN in medical imaging applications is the FRRnet\cite{RN20}. The network has two pathways: a full resolution path that helps in identifying precise boundaries and a multi-scale feature extraction pathway that is responsible for robust feature recognition. Also, the residual connections improve the gradient flow through the network\cite{RN58}. By combining the information from both the pathways, the FRRnet was able achieve precise localization and robust feature recognition\cite{RN20}. \\

Many studies have demonstrated that an ensemble network that learned to combine the predictions of multiple DCNNs into a single predictive model offered a better accuracy than each of the networks separately\cite{RN59,RN60}. When trained on the same training data as the individual DCNNs (weights of the individual DCNNs were frozen), the ensemble network learned to reduce the variance for each network, thus dramatically increasing the predictive power. \\

\begin{figure}[H]
    \centering
    \includegraphics[scale=0.85]{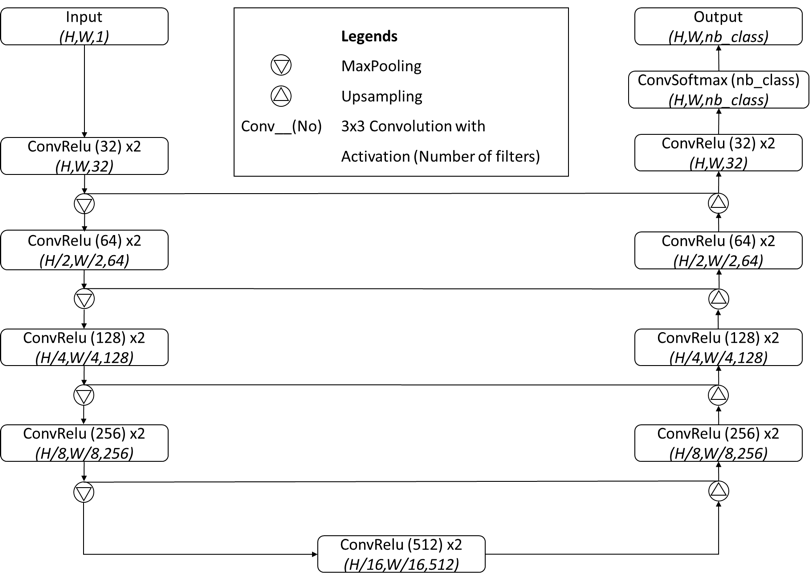}
    \caption{U-net architecture.}
    \label{suppli:1}
\end{figure}

\begin{figure}[H]
    \centering
    \includegraphics[scale=0.85]{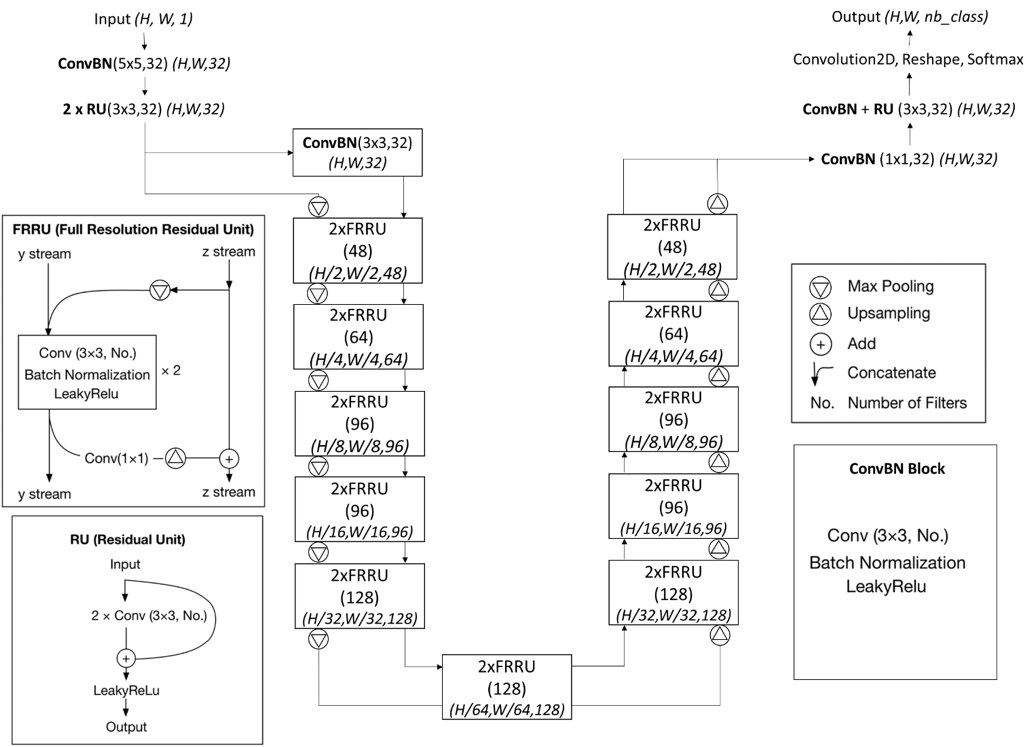}
    \caption{FRRN architecture.}
    \label{suppli:2}
\end{figure}

\begin{figure}[p]
    \centering
    \includegraphics[scale=1.0]{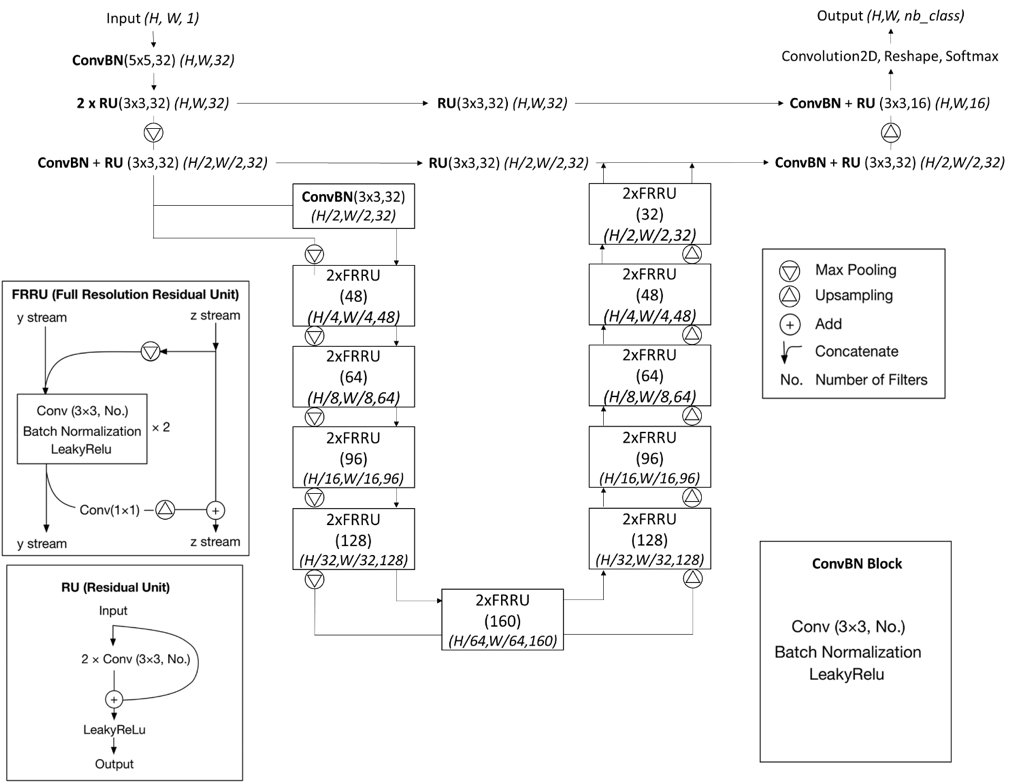}
    \caption{Proposed Network: Hybrid between U-net and Full Resolution Residual Net – FRRUnet.}
    \label{suppli:3}
\end{figure}

\begin{figure}[H]
    \centering
    \includegraphics[scale=0.85]{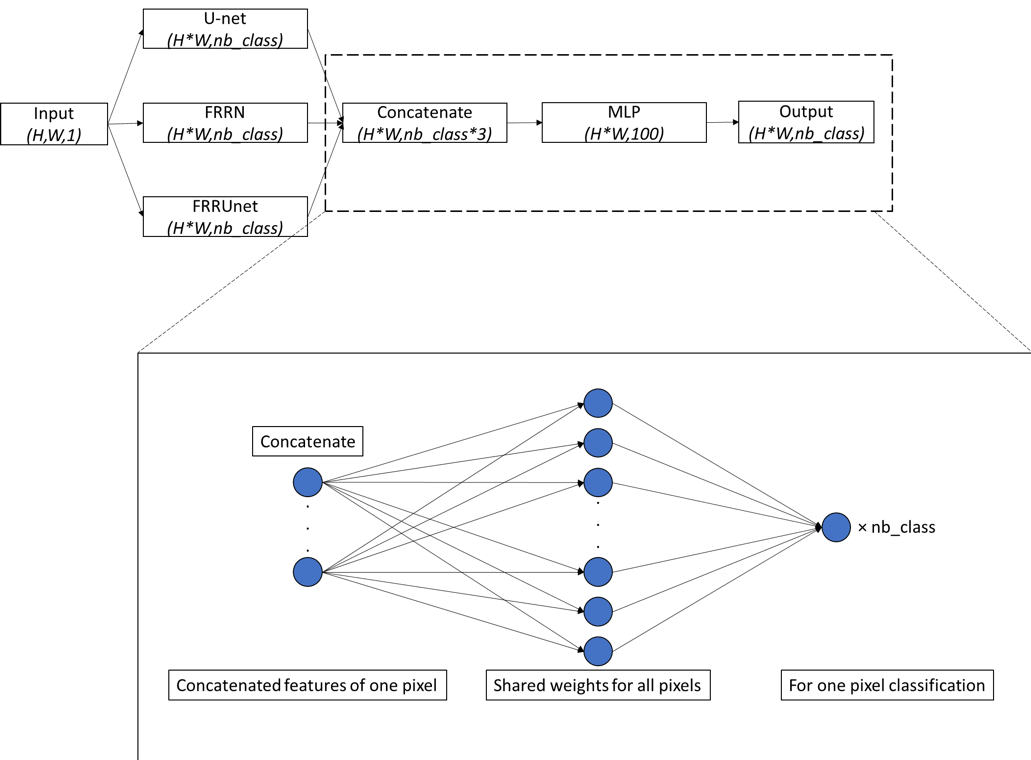}
    \caption{Ensemble architecture that combine the three base models.}
    \label{suppli:4}
\end{figure}

In this study, we developed FRRUnet (full resolution residual U-Net), a hybrid DCNN that exploited the inherent advantages of both the U-Net and the FRRnet. For the detection of the SSL, the FRRUnet was used, while an ensemble of the U-Net, FRRnet, and the FRRUnet was used for the segmentation of the ASOCT structures [\textbf{Figure \ref{suppli:1}}, \textbf{Figure \ref{suppli:2}}, \textbf{Figure \ref{suppli:3}}, \textbf{Figure \ref{suppli:4}}]. \\

All three networks were trained end to end using an Adam optimizer and categorical cross entropy loss function\cite{RN6} with a learning rate of 0.00005. All the convolution layers were activated with a leaky rectifier linear unit (ReLU)\cite{RN21} activation function. A dropout layer with a probability of 0.5 was used after every building block to reduce the overfitting. Given the limited size of the training dataset, the DCNNs’ variance was increased through data augmentation techniques such as rotation, width shift, height shift, shear, zoom, flip, brightness and contrast shift. The final U-Net, FRRnet, FRRUnet, and the ensemble network consisted of 7.80 M, 4.2 M, 4.2 M, and 1.7K trainable parameters respectively. All networks were trained and tested on an NVIDIA GTX 1080 founder’s edition GPU with CUDA v8.0 and cuDNN v5.1 acceleration. Using the given hardware configuration, for each ASOCT image the network was able to detect the SSL in 0.108 $\pm$ 0.0035 seconds and segment the ASOCT tissues in 0.324 $\pm$ 0.0018 seconds. The parameters were then automatically computed on a CPU (Intel Xeon at 2.1 GHz) in under 1.723 $\pm$ 0.287 seconds. It should be noted that parameter measurement can be accelerated by parallelism since each scan is independent. \\


\begin{table}[H]
\centering
\caption {\textbf {Definitions of important anterior segment optical coherence tomography parameters.}} 
\begin{tabular}{@{}ll@{}}
\toprule
\textbf{Parameter}                & \textbf{Definition}                                                                                                                                                                                                                                                                        \\ \midrule
Anterior Chamber Depth (ACD)      & Axial distance between corneal endothelium to anterior lens surface\cite{RN40}                                                                                                                                                                                                                       \\ \midrule
Lens Vault (LV)                   & \begin{tabular}[c]{@{}l@{}}Perpendicular distance from middle of the line connecting the scleral\\ spurs to the anterior pole of the lens\cite{RN39}\end{tabular}                                                                                                                                     \\ \midrule
Anterior Chamber Width (ACW)      & Distance between two scleral spurs\cite{RN41}                                                                                                                                                                                                                                                         \\ \midrule
Anterior Chamber Area (ACA)       & \begin{tabular}[c]{@{}l@{}}Area bordered by posterior surface of the cornea, anterior surface of \\ iris and anterior surface of the lens\cite{RN42}\end{tabular}                                                                                                                                     \\ \midrule
Angle Opening Distance (AOD)      & \begin{tabular}[c]{@{}l@{}}Distance between the anterior iris surface and posterior corneal surface\\ on a line perpendicular to the trabecular meshwork, a distance from \\ the scleral spur (500 $\mu$m, 750 $\mu$m, etc.)\cite{RN43}\end{tabular}                                                            \\ \midrule
Trabecular Iris Space Area (TISA) & \begin{tabular}[c]{@{}l@{}}Area of a trapezoid created by the following boundaries: AOD of a \\ distance from scleral spur (500 $\mu$m, 750 $\mu$m, etc.), \\ line from scleral spur perpendicular to plane of inner scleral wall to\\ the iris, inner corneoscleral wall, iris surface\cite{RN43}\end{tabular} \\ \midrule
Iris thickness (IT)               & \begin{tabular}[c]{@{}l@{}}IT at a distance from the scleral spur or a relative distance \\ in the iris (eg: middle of iris)\cite{RN44}\end{tabular}                                                                                                                                                  \\ \midrule
Iris Curvature                    & \begin{tabular}[c]{@{}l@{}}Distance from iris greatest convexity point to the line between \\ most central and most peripheral iris pigment epithelium\cite{RN44}\end{tabular}                                         
                                                                            \\ \bottomrule
\label{tab:table1}
\end{tabular}
\end{table}


\subsection{Inter and intra observer tests}
We performed an inter-observer agreement test to assess the reproducibility when identifying the scleral spur between three human observers: A – Trained non-expert, B – Trained medical student; C – Fellowship-trained glaucoma expert well-versed in ASOCT analysis and the software algorithm. The intra-observer agreement test assessed the extent of repeatability among the human observers and their comparison with the software algorithm. A paired t-test was used to measure the extent of agreement on-average and Bland-Altman plots were used to depict the limit of agreement ($\pm$1.96 SD) and the distribution of discrepancy between individual measurements. The intra-correlation coefficient (ICC), assessed with a single rater, absolute agreement, two-way random effect model, was used to reflect the degree of agreement and correlation between measurements. ICCs of $<$0.50, 0.50-0.75, 0.75-0.90; $<$0.90 were taken as poor, moderate, good and excellent measures of reliability, respectively. All p-values presented are 2-sided and statistically significant if $<$0.05. \\

\subsection{Quality check}
Poor quality scans (low signal strength, presence of motion/blink artefact, improper head positioning etc.) can affect the localization and segmentation performance of the DCNNs, thus resulting in incorrect automated measurements. In this study, we performed a two-step automated quality check based on the predictions obtained to eliminate poor quality ASOCT images. First, upon the detection of the SSL a square region surrounding the center of the predicted region was obtained as the reference. A confidence index was computed as the intersection over union (IoU; between 0-1) between the predicted and reference regions. Scans that yielded a confidence index greater than or equal to 0.80 were considered good, while lower values were designated as poor quality. Second, for the segmentation the number of closed and continuous contours representing each class were used to assess the quality of a scan, i.e., the iris should have two contours, while the corneo-sclera shell and the anterior chamber should have only a single contour each. Scans with predictions that did not satisfy these criteria were considered as poor quality. Finally, the automatically extracted parameters were considered reliable only if the ASOCT scan satisfied both the aforementioned quality check criteria. The test images are made sure to be of usable quality clinically.

\section{Results}

All results in this section are from 4 observers: A -- trained non-expert, B -- trained medical student, C -- fellowship-trained glaucoma expert well-versed in ASOCT analysis, M -- Trained machine. The same denotation is used throughout. The mean age of the patients was 62.20 $\pm$ 8.35 and 31.91\% of them were males. The percentage for Chinese, Malay, Indian and other races was 77.86\%, 11.42\%, 7.86\% and 2.86\% respectively.

\subsection{Scleral spur localization}
\begin{figure}[!pt]
    \centering
    \includegraphics[scale=0.80]{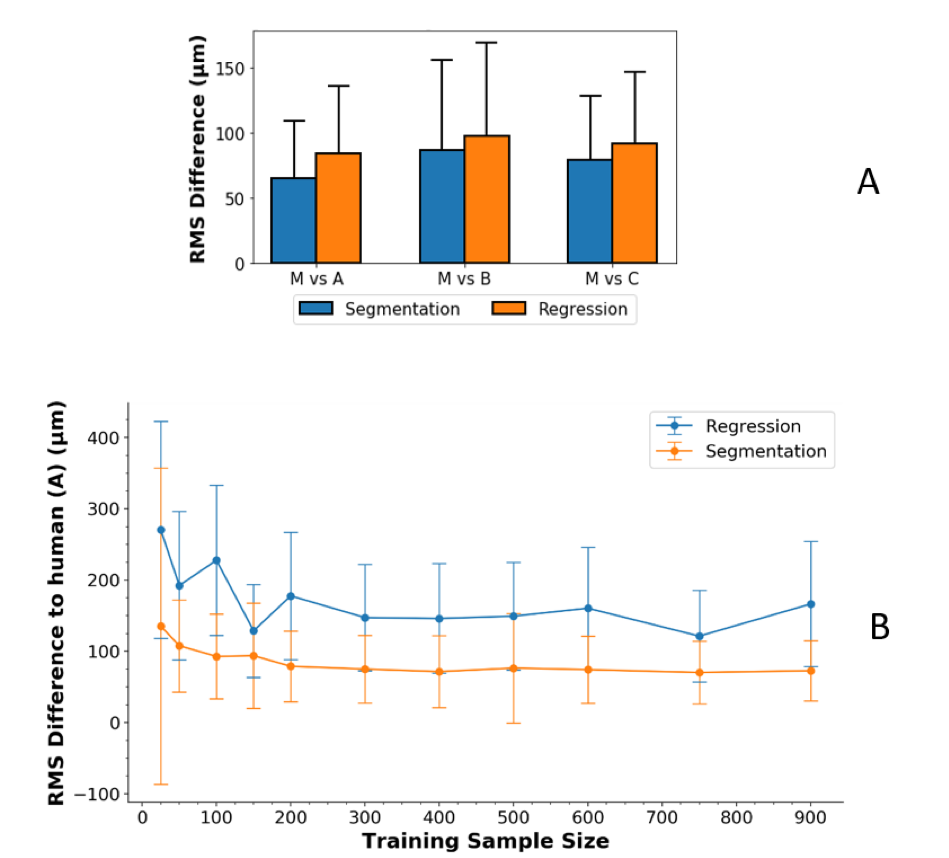}
    \caption{Segmentation vs Regression Approach. \textbf{A}: Inter-observer test against human observers. \textbf{B}: Varying training sample size and calculated distance against human observer.}
    \label{fig:1}
\end{figure}

First, our proposed segmentation approach was compared against a regression approach, both utilizing DCNNs. The final models were trained for 1,000 iterations and then tested against 3 human observers \textbf{Figure \ref{fig:1} A}. The segmentation approach was closer to human observers for all cases. The next test showed that our segmentation approach could reach human level detection with a much smaller training dataset ($\sim$200 samples or $\sim$100 images) \textbf{Figure \ref{fig:1} B}.\\

\begin{figure}[!pb]
    \centering
    \includegraphics[scale=0.80]{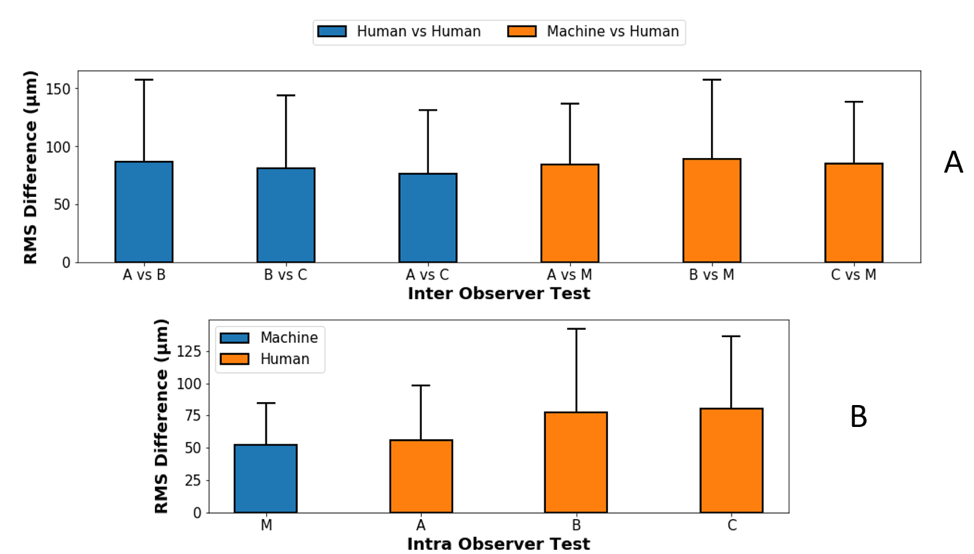}
    \caption{Observer Test results. \textbf{A}: Inter-observer Test. \textbf{B}: Intra-observer Test.}
    \label{fig:2}
\end{figure}

\begin{figure}[p]
    \centering
    \includegraphics{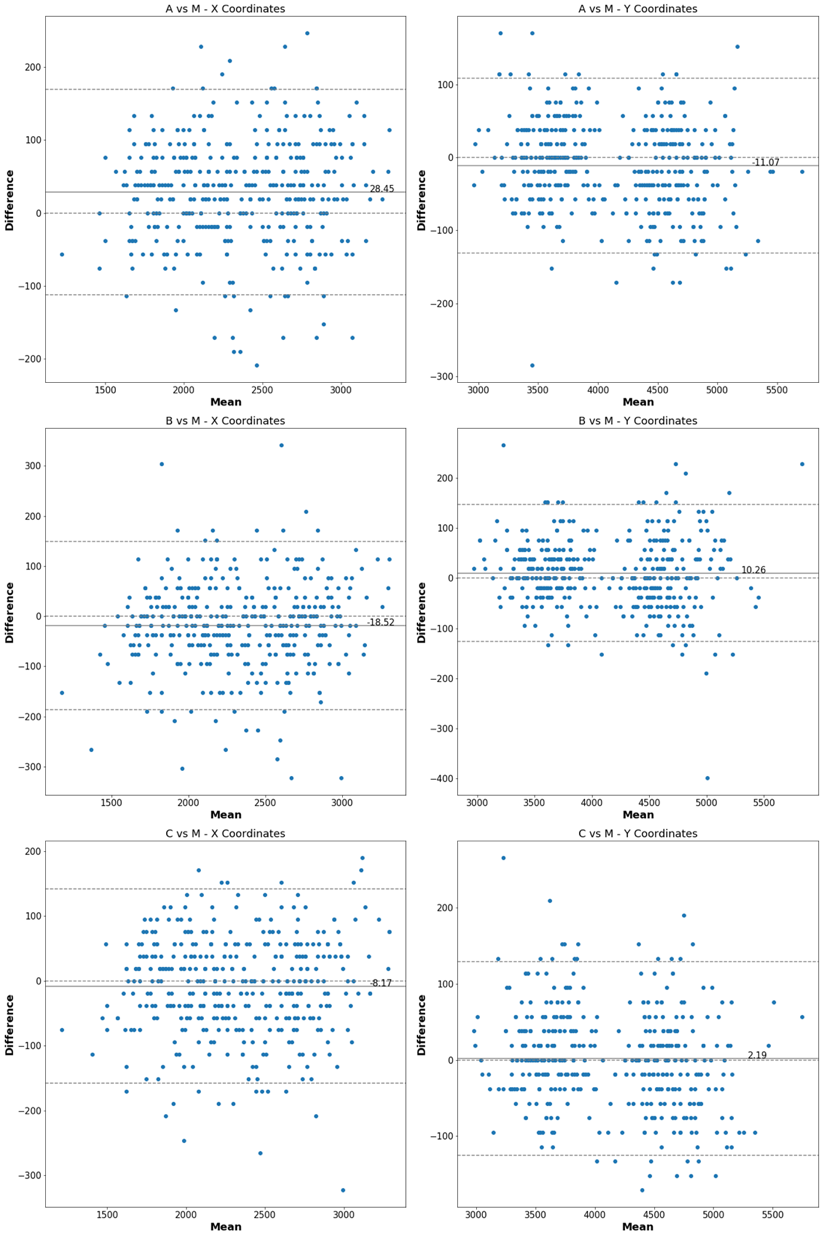}
    \caption{Bland-Altman Plot for Human-Machine Inter Observer Test, both axes measured in $\mu$m.}
    \label{suppli:6}
\end{figure}

Inter-observer tests showed that human grader differences were not significantly different from human and machine differences in most cases \textbf{Figure \ref{fig:2} A}. Moreover, intraclass correlation (ICC) was done for each observer pair for the X and Y coordinates of the scleral spur location \textbf{Table \ref{tab:table2}}.  It was shown that the machine’s scleral spur marking was in high agreement with human graders. Bland-Altman plots for Machine -- Human pair was further provided in \textbf{Figure \ref{suppli:6}}.\\

\begin{table}[H]

\centering
\caption {\textbf {ICC results for Inter Observer Test}}
\begin{tabular}{@{}lllllllllll@{}}
\toprule
\multicolumn{11}{c}{\textbf{Two-way Single Score Absolute Agreement ICC}}                                                                                                                                                                                                                                               \\ \midrule
\multicolumn{1}{|l|}{X Coordinate} & \multicolumn{1}{l|}{A} & \multicolumn{1}{l|}{B} & \multicolumn{1}{l|}{C} & \multicolumn{1}{l|}{M} & \multicolumn{1}{l|}{\multirow{5}{*}{}} & \multicolumn{1}{l|}{Y Coordinate} & \multicolumn{1}{l|}{A} & \multicolumn{1}{l|}{B} & \multicolumn{1}{l|}{C} & \multicolumn{1}{l|}{M} \\ \cmidrule(r){1-5} \cmidrule(l){7-11} 
\multicolumn{1}{|l|}{A}            & 1                      & 0.978                  & 0.985                  & 0.984                  & \multicolumn{1}{l|}{}                  & \multicolumn{1}{l|}{A}            & 1                      & 0.993                  & 0.995                  & 0.994                  \\ \cmidrule(r){1-5} \cmidrule(l){7-11} 
\multicolumn{1}{|l|}{B}            &                        & 1                      & 0.983                  & 0.979                  & \multicolumn{1}{l|}{}                  & \multicolumn{1}{l|}{B}            &                        & 1                      & 0.994                  & 0.993                  \\ \cmidrule(r){1-5} \cmidrule(l){7-11} 
\multicolumn{1}{|l|}{C}            &                        &                        & 1                      & 0.984                  & \multicolumn{1}{l|}{}                  & \multicolumn{1}{l|}{C}            &                        &                        & 1                      & 0.993                  \\ \cmidrule(r){1-5} \cmidrule(l){7-11} 
\multicolumn{1}{|l|}{M}            &                        &                        &                        & 1                      & \multicolumn{1}{l|}{}                  & \multicolumn{1}{l|}{M}            &                        &                        &                        & 1                      \\ \bottomrule
\label{tab:table2}

\end{tabular}
\end{table}

The machine neural network was deterministic once training was complete, meaning that a given input always resulted in the same output. Hence, to do intra-observer tests, another model was trained from scratch and used to compare with the first model. RMS difference for the machine intra-observer test was significantly smaller than most human intra-observer tests (except for observer A, whose intra-observer result was similar to the machine)(\textbf{Figure \ref{fig:2} B}).  This means that machine SSL prediction generally had lower variability than that of human grader.\\

\subsection{ASOCT segmentation}

\begin{figure}[H]
    \centering
    \includegraphics[scale=0.90]{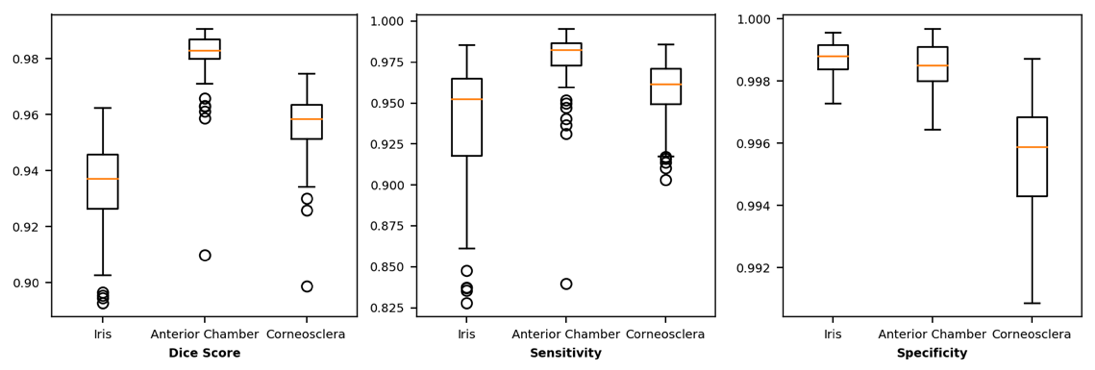}
    \caption{Validation scores for ASOCT segmentation.}
    \label{fig:3}
\end{figure}

The ASOCT segmentation performance of the trained network was validated using the Dice coefficient, sensitivity and specificity \textbf{Figure \ref{fig:3}}, as described below. The Dice coefficient was used to assess the similarity between the manual segmentation and DCNN segmentation. The coefficient was defined between 0 and 1 (0: no overlap; 1: perfect overlap), and was calculated for each class as follows:

\begin{align*}
\textrm{Dice score}= \frac{2 \times |D \cap M|}
{2 \times |D \cap M| + |D \setminus M| + |M \setminus D|}
\end{align*}

where D and M are the set of pixels representing the particular class in the DCNN and manual segmentation, respectively.

Specificity and sensitivity were used to obtain the true negative (assess false predictions) and true positive rates (assess correct predictions) respectively. They were defined for each class as follows: 

\begin{align*}
\textrm{Specificity}= \frac{|\overline{D} \cap \overline{M}|}
{|\overline{M}|}
\end{align*}

\begin{align*}
\textrm{Sensitivity}= \frac{|{D} \cap {M}|}
{|{M}|}
\end{align*}

\begin{figure}[p]
    \centering
    \includegraphics[scale=0.9]{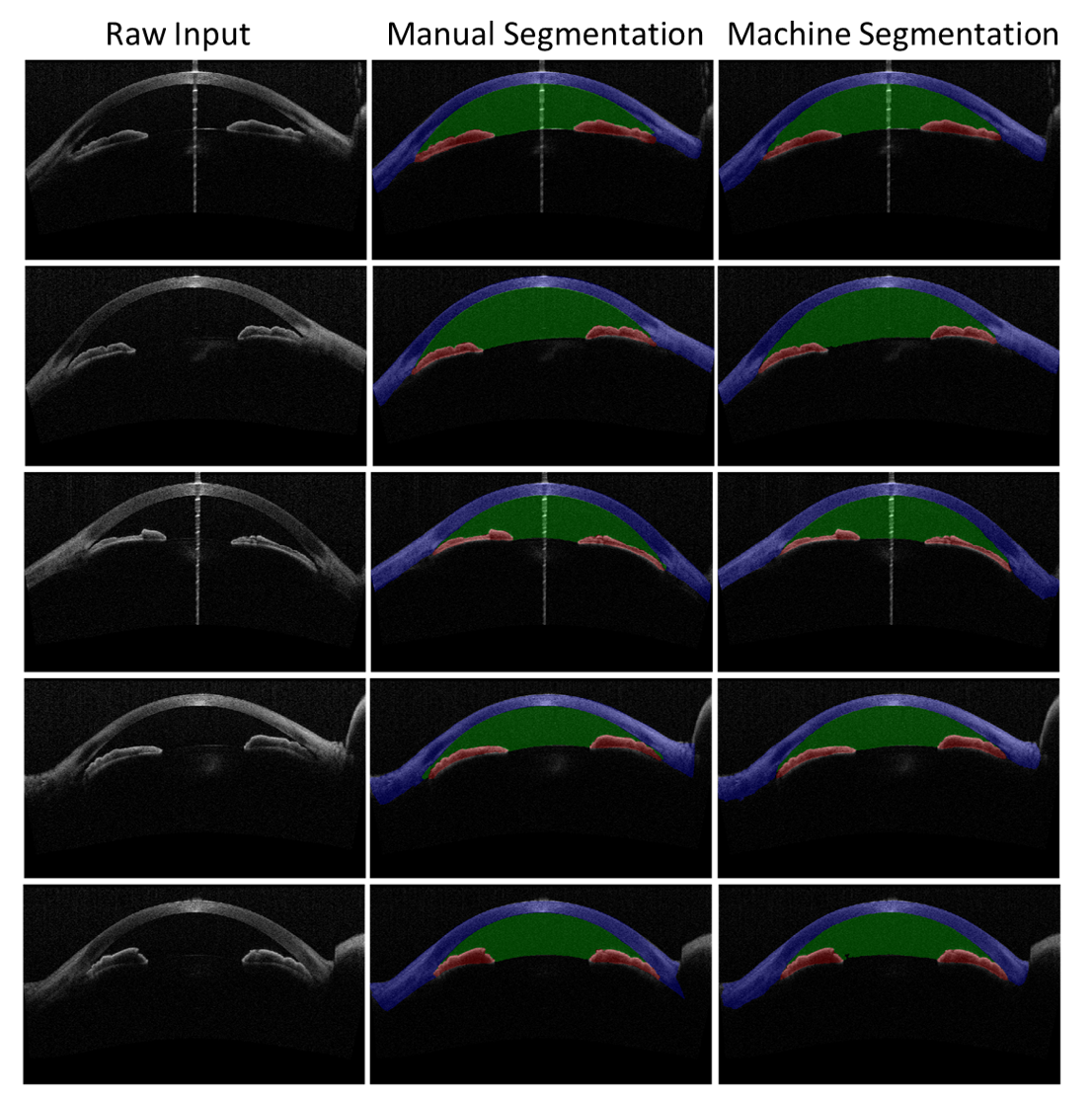}
    \caption{Example predictions on test set versus human manual segmentation.}
    \label{suppli:5}
\end{figure}

Both specificity and sensitivity were defined between 0 and 1.
Examples of machine segmentation results can be found in \textbf{Figure \ref{suppli:5}}.\\

\clearpage
\subsection{Parameter extraction}

\begin{figure}[H]
    \centering
    \includegraphics{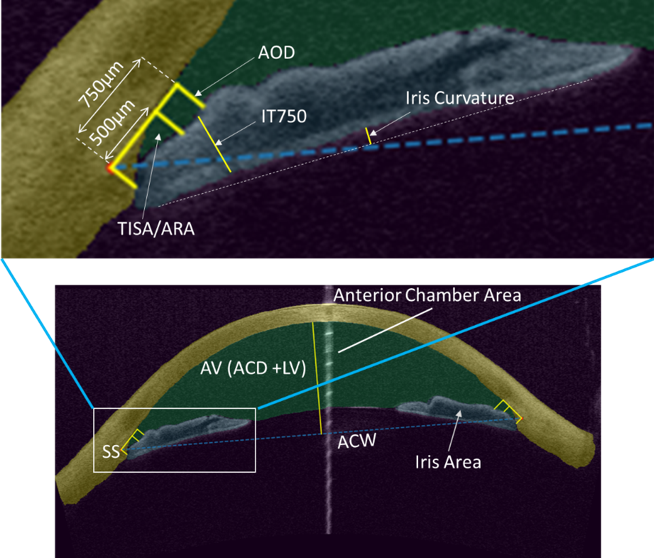}
    \caption{ASOCT Parameter Extraction and Definitions. \textbf{Anterior Chamber Depth (ACD)}: axial distance between corneal endothelium to anterior lens surface. \textbf{Lens Vault (LV)}: perpendicular distance from the middle of the line connecting the scleral spurs to the anterior pole of the lens. \textbf{Anterior Chamber Width (ACW)}: distance between the two scleral spurs. \textbf{Anterior Chamber Area (ACA)}: the area bordered by posterior surface of the cornea, anterior surface of iris and anterior surface of the lens. \textbf{Angle Opening Distance (AOD)}: distance between the anterior iris surface and posterior corneal surface on a line perpendicular to the trabecular meshwork, at a specific distance from the scleral spur (500 $\mu$m, 750 $\mu$m etc.). \textbf{Trabecular Iris Space Area (TISA)}: area of a trapezoid created by the following boudaries: AOD of a distance from scleral spur (500 $\mu$m, 750 $\mu$m etc.), line from scleral spur perpendicular to plane of inner scleral wall to the iris, inner corneoscleral wall, iris surface. \textbf{Iris thickness (IT)}: IT at a distance from the scleral spur or a relative distance in the iris (eg: middle of iris). \textbf{Iris Curvature (ICurve)}: distance from iris greatest convexity point to the line between most central and most peripheral iris pigment epithelium.}
    \label{fig:4}
\end{figure}

Parameter extraction was a crucial step to help validate the scleral spur localization. The segmentation used in this step was fully automated, based on the assumption that the accuracy of automated ASOCT segmentation is already high. \textbf{Figure \ref{fig:4}} defined the measured ACA parameters. \textbf{Table \ref{tab:table3}} shows ICC results for inter- and intra-observer test agreement. Inter-observer test results showed good to excellent agreement between observers, especially between machine and human. Moreover, for measurements with relatively lower ICC between machine and human, the human-human counterpart results were similar.  Intra-observer test ICC for machine was higher than human, indicating that the machine was more consistent and stable.\\

\begin{table}[H]
\centering
\caption {\textbf {ICC results for Inter and Intra Observer Tests for ASOCT parameter extraction for ACW, TISA and AOD}}

\begin{tabular}{@{}lllll@{}}
\multicolumn{5}{c}{\textbf{Inter Observer Test (Two-way, single score, absolute agreement ICC)}}                                                                                                        \\ \cmidrule(l){2-5} 
\multicolumn{1}{l|}{}                  & \multicolumn{1}{l|}{\textbf{A vs M}} & \multicolumn{1}{l|}{\textbf{B vs M}} & \multicolumn{1}{l|}{\textbf{C vs M}} & \multicolumn{1}{l|}{\textbf{A vs B vs C}} \\ \midrule
\multicolumn{1}{|l|}{\textbf{ACW}}     & 0.941                                & 0.931                                & 0.949                                & 0.937                                     \\ \cmidrule(r){1-1}
\multicolumn{1}{|l|}{\textbf{TISA500}} & 0.784                                & 0.722                                & 0.710                                & 0.759                                     \\ \cmidrule(r){1-1}
\multicolumn{1}{|l|}{\textbf{TISA750}} & 0.822                                & 0.728                                & 0.761                                & 0.793                                     \\ \cmidrule(r){1-1}
\multicolumn{1}{|l|}{\textbf{AOD500}}  & 0.910                                & 0.902                                & 0.927                                & 0.926                                     \\ \cmidrule(r){1-1}
\multicolumn{1}{|l|}{\textbf{AOD750}}  & 0.880                                & 0.863                                & 0.898                                & 0.903                                     \\ \cmidrule(r){1-1}
\multicolumn{5}{l}{\textbf{Intra Observer Test (Two-way, single score, absolute agreement ICC)}}                                                                                                        \\ \cmidrule(l){2-5} 
\multicolumn{1}{l|}{}                  & \multicolumn{1}{l|}{\textbf{M}}      & \multicolumn{1}{l|}{\textbf{A}}      & \multicolumn{1}{l|}{\textbf{B}}      & \multicolumn{1}{l|}{\textbf{C}}           \\ \midrule
\multicolumn{1}{|l|}{\textbf{ACW}}     & 0.979                                & 0.951                                & 0.953                                & 0.954                                     \\ \cmidrule(r){1-1}
\multicolumn{1}{|l|}{\textbf{TISA500}} & 0.847                                & 0.845                                & 0.728                                & 0.646                                     \\ \cmidrule(r){1-1}
\multicolumn{1}{|l|}{\textbf{TISA750}} & 0.884                                & 0.887                                & 0.738                                & 0.702                                     \\ \cmidrule(r){1-1}
\multicolumn{1}{|l|}{\textbf{AOD500}}  & 0.959                                & 0.958                                & 0.923                                & 0.881                                     \\ \cmidrule(r){1-1}
\multicolumn{1}{|l|}{\textbf{AOD750}}  & 0.948                                & 0.956                                & 0.874                                & 0.901                                     \\ \cmidrule(r){1-1}  

\label{tab:table3}

\end{tabular}
\end{table}

\subsection{Results visualization and quality check}

\begin{figure}[H]
    \centering
    \includegraphics[scale=1]{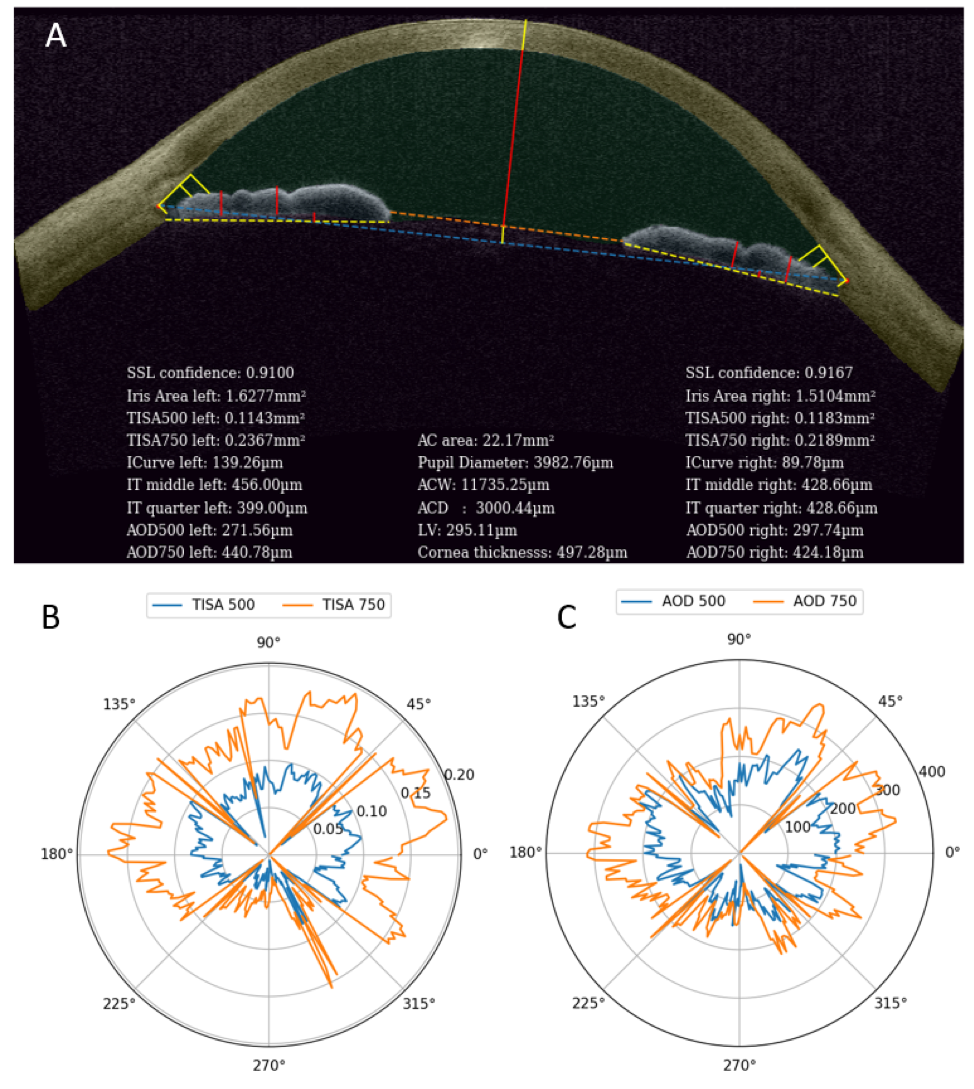}
    \caption{Example of automated results. \textbf{(A)} Example parameter measurement of a single scan. \textbf{(B)} Example of 360$^{\circ}$ analysis for AOD. \textbf{(C)} Example 360$^{\circ}$ analysis for TISA. The measured value for each scan in the whole volume is denoted by the radius, while the angle corresponds to the scan position in the ASOCT volume.}
    \label{fig:5}
\end{figure}

This was assessed visually by exporting the software prediction into an image format. The machine was able to visualize the per-scan results \textbf{Figure \ref{fig:5} A}. Moreover, fully automated measurement enables 360$^{\circ}$ analysis, for example of AOD and TISA \textbf{Figure \ref{fig:5} B and C}. The goniogram showed that the inferior quadrant’s angle is narrower than other quadrants of that specific patient\textquotesingle s eye \textbf{Figure \ref{fig:5} B and C}. Indicating that a global assessment would provide a more accurate diagnosis.\\

\begin{figure}[H]
    \centering
    \includegraphics[scale=1]{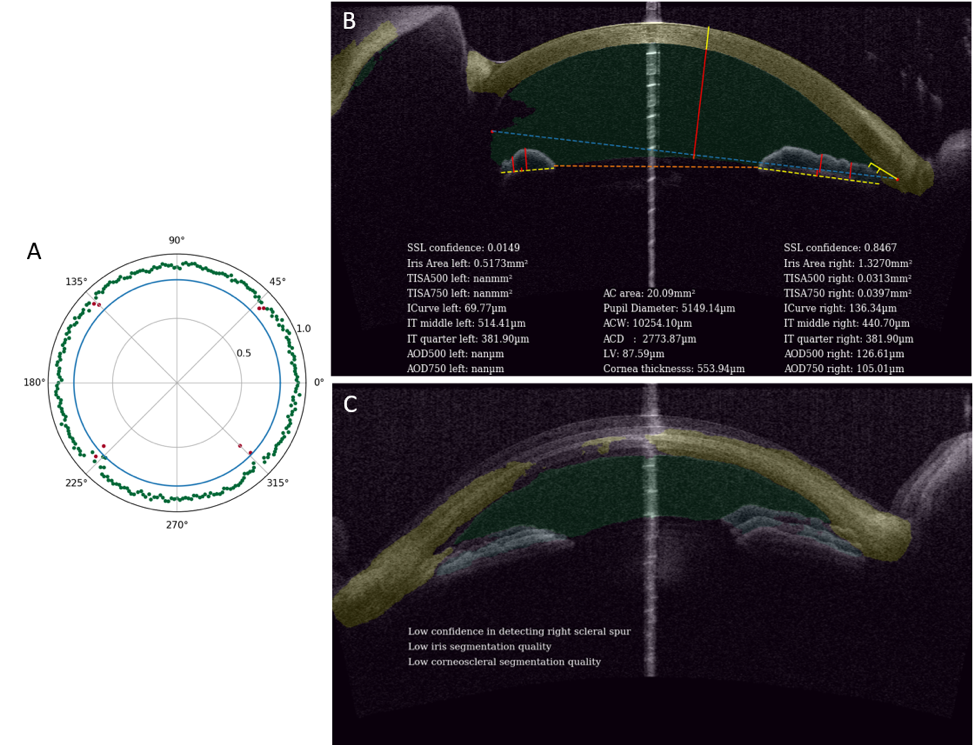}
    \caption{Example of quality check results. \textbf{(A)} Visualization of SSL confidence 360$^{\circ}$. Greens are passed scans. Reds are failed scans. Blue circle is 0.8 SSL confidence threshold. Red dots above the thresholds are scans that failed the ASOCT segmentation check. In this example 4/128 scans are disqualified. \textbf{(B)} Excluded scan due to low SSL confidence on the left side. \textbf{(C)} Excluded scan due to bad segmentation quality.}
    \label{fig:6}
\end{figure}

For image quality check, the ASOCT scans need to pass both the SSL confidence and ASOCT segmentation quality assessment. The SSL confidence can be visualized in 360$^{\circ}$ as shown in Figure \textbf{Figure \ref{fig:6} A}. Visually comparison of good \textbf{Figure \ref{fig:5} A} and failed \textbf{Figure \ref{fig:6} B and C} cases determined that, if the image quality is good, the SSL confidence should be above 0.85. Accordingly, the SSL confidence threshold was set to 0.8, in order to have some margin from 0.85, meaning scans with SSL confidence below 0.8 were excluded. Moreover, this threshold can be manually adjusted. A failed SSL detection can be seen in \textbf{Figure \ref{fig:6} B} on the left scleral spur, where SSL confidence is accordingly very low. For ASOCT segmentation, the exclusion criteria are for iris, anterior chamber, corneo-sclera, a number of contours larger than 5, 6 and 10, respectively. Ideally, the number of contours for the said areas of interest should be 2, 1 and 1 respectively. However, for narrow angle cases and many other noisy cases, there might be insignificant wrong small contours. Hence, we increased the threshold. All of these are hyper-parameters and can be tuned. A future systematic study of hyper-parameter tuning is planned. A failed ASOCT segmentation can be seen in \textbf{Figure \ref{fig:6} C}. All failed scans were excluded from the final parameter extraction.\\

\section{Discussion}

The use of ASOCT for the assessment of the ACA in angle closure glaucoma is increasingly popular in the clinical setting. However, the practicality and efficiency of its assessment remains challenging for the ophthalmologists. In the absence of an absolute ground truth for SSL, any prediction, including that of experienced human graders, may be expected to contain errors and show variability in performance. The errors consist of bias, variance and irreducible error (noise)\cite{RN13,RN8}. Thus, when a machine learns from human graders, it also learns the human’s error. However, with more trainers and data, the errors would be centered around zero\cite{RN13,RN62}. In addition, if the algorithm is developed using expert trainers’ inputs, these errors would stabilize faster. In clinical practice, errors and variability in SSL on ASOCT scans have huge impact in the diagnosis of angle closure glaucoma because incorrect identification of SSL can result in misdiagnosis and management of patients with PACG. ASOCT imaging has been shown to be more objective and quantifiable compared to gonioscopic techniques\cite{RN11,RN38,RN62,RN63}. The ACA measurements from ASOCT scans are heavily dependent on the SSL and ophthalmologists gauge treatment effectiveness based on ASOCT measurements before and after treatment. \\

One of the strengths of the presented method is that it utilizes 3 different approaches to identify the SSL, allowing the machine to be more robust and, thus, be able to more accurately locate the SSL on a variety of ASOCT scans. For ASOCT segmentation, beside a high Dice coefficient, the network also had high sensitivity and specificity, making it a reliable tool in quantifying ASOCT parameters. A comparable algorithm is the STAR Program available on the Casia 2 swept-source ASOCT (Tomey Corporation, Nagoya, Japan), which is capable of automated identification of SSL and ACA measurements\cite{RN53}.However, this program is a semi-automated software which uses simple edge detection to detect the scleral spur-uvea edge line and, from that, detect the scleral spur location\cite{RN53}. Moreover, it also depends on the assumption that SSL lies in a perfect circle. In cases of narrow angle, there will be iridotrabecular contact and the scleral spur-uvea edge line will not be visible. In our approach, the machine is trying to learn from human expertise, hence it can detect the scleral spur without the edge line and it also has the potential to expand its definition of scleral spur implicitly by learning from the expert human grader.\\

The main limitation in our study was the lack of an absolute ground truth in labelling of the ASOCT images. The labelled data was being prepared by human trainers. This is compounded by crowding of the ACA in eyes with angle closure. The compressed ocular tissues, namely the cornea, peripheral iris and trabecular meshwork, make accurate identification of the scleral spur challenging. Hence, one of the limitations of the paper is the lack of trainers. To validate the machine’s performance without a true ground truth, we used the inter- and intra-observer test and ICC, with the exception of the ASOCT segmentation where we only had one trainer and observer. Through the validation tests conducted, it was shown that the machine performance was in good agreement with human performance, while the former was more consistent.\\

As mentioned before, the lack of a generalized population of trainers caused the machine’s performance to be biased towards the trainers’ errors. As shown in our inter-observer test, since observer A was a trainer for the network, the distance between machine and observer A was lower than the machine with observer B or C. This limitation could be resolved simply by having more trainers. The second limitation was the presence of only one expert. Again, this could be resolve by having more experts.\\

One technical limitation of our approach was that the resolution depends on the Focus region. The landmarks could not lie too close to the border. The distance should be larger than half of the focus region length, since the point of interest lay in the center of the region. This could be resolved partially with padding (introduce non-meaningful features) or decreasing the size of focus region (susceptible to class imbalances\cite{RN15}).\\

The impact of our method of accurate and automated identification of the scleral spur in ASOCT scans would be in the diagnosis and monitoring of angle closure glaucoma eyes. The diagnosis of angle closure on ASOCT images is dependent on accurate localization of the scleral spur. Angle closure is defined by contact between the peripheral iris and the trabecular meshwork anterior to the scleral spur\cite{RN11}. As such, the accurate localization of the scleral spur can potentially make screening of angle closure glaucoma on ASOCT imaging easier and more automated. This is especially useful for modern swept-source ASOCT which provides a 360-degree scan of the eye and as many as 64 cross-section cuts of the ACA per eye. The automated identification of the scleral spur reduces variability of human graders and speeds up image analysis to provide a more comprehensive evaluation of the ACA. In the monitoring of angle closure glaucoma eyes, the ACA characteristics should be tracked over time and this paper demonstrates how these parameters can be measured in a reproducible manner, as most ACA measurements use the scleral spur as the reference. These ACA parameters are important in determining the mechanisms of angle closure, guiding clinical management and measuring efficacy of treatment modalities\cite{RN55,RN56}.

\section*{Disclosures}
The authors declare that there are no conflicts of interest related to this article.

\bibliographystyle{plain}
\bibliography{asoct}

\end{document}